\documentclass[twocolumn,aps,prl,preprintnumbers,superscriptaddress]{revtex4}
\usepackage{}
\usepackage{bbm}
\usepackage[latin9]{inputenc}
\usepackage{amsmath}
\usepackage{amssymb}
\usepackage{graphicx}
\usepackage{mathrsfs}
\usepackage{amsfonts}
\usepackage{amsthm}
\usepackage{color}
\usepackage{txfonts}
\usepackage{lipsum}
\usepackage{ulem}
\usepackage{gensymb}
\usepackage[colorlinks=true,citecolor=blue,linkcolor=blue,urlcolor=blue,anchorcolor=blue]{hyperref}%
\hypersetup{colorlinks=true,citecolor=blue,linkcolor=blue,urlcolor=blue}
\providecommand{\U}[1]{\protect\rule{.1in}{.1in}}
\makeatletter

\newcommand{\Rmnum}[1]{\expandafter\@slowromancap\romannumeral #1@}
\makeatother
\makeatletter
\@ifundefined{textcolor}{}
{
\definecolor{BLACK}{gray}{0}
\definecolor{WHITE}{gray}{1}
\definecolor{RED}{rgb}{1,0,0}
\definecolor{GREEN}{rgb}{0,1,0}
\definecolor{BLUE}{rgb}{0,0,1}
\definecolor{CYAN}{cmyk}{1,0,0,0}
\definecolor{MAGENTA}{cmyk}{0,1,0,0}
\definecolor{YELLOW}{cmyk}{0,0,1,0}
}
\makeatother
\begin{document}
\title{Peculiar corner states in magnetic fractals}
\author{Zhixiong Li}
%\email[Corresponding author: ]{zhixiong\_li@csu.edu.cn}
\affiliation{School of Physics, Central South University, Changsha 410083, China}
\author{Peng Yan}
\email[Corresponding author: ]{yan@uestc.edu.cn}
\affiliation{School of Physics and State Key Laboratory of Electronic Thin Films and Integrated Devices, University of Electronic Science and Technology of China, Chengdu 610054, China}

\begin{abstract}
Topological excitations in periodic magnetic crystals have received significant recent attention. However, it is an open question on their fate once the lattice periodicity is broken. In this work, we theoretically study the topological properties embedded in the collective dynamics of magnetic texture array arranged into a Sierpi\'nski carpet structure with effective Hausdorff dimensionality $d_{f}=1.893$. By evaluating the quantized real-space quadrupole moment, we obtain the phase diagram supporting peculiar corner states that are absent in conventional square lattices. We identify three different higher-order topological states, i.e., outer corner state, type I and type II inner corner states. We further show that all these corner states are topologically protected and are robust against moderate disorder. Full micromagnetic simulations are performed to verify theoretical predictions with good agreement. Our results pave the way to investigating topological phases of magnetic texture based fractals and bridging the gap between magnetic topology and fractality.   
\end{abstract}

\maketitle
\textit{Introduction}.$-$Due to the potential applications of topological insulators (TIs) in information transmission and quantum computing \cite{HasanRMP2010,QiRMP2011,Libook2023}, searching the topological phase of matter has become one of the central topics in physics and engineerings. Moreover, the discovery of higher-order topological insulators (HOTIs) \cite{BenalcazarS2017,SchindlerSA2018,HassanNP2019,XueNM2019,ImhofNP2018,FanPRL2019,Linpj2019} has further extended the scope of topology family. In condensed matter physics, the natural and artificial materials are generally characterized in the context of crystals which consist of periodically arranged atoms with the translational symmetry. The quantized topological invariants in momentum space can therefore be conveniently computed to describe the topological states, which constitutes the so-called ``bulk-boundary" or ``bulk-corner" correspondence \cite{HatsugaiPRL1993,BenalcazarPRB2017,XieNRP2021,LiPR2021}. Interestingly, there also exist some other materials which have ordered structure but do not support translational symmetry, such as the quasicrystal \cite{BindiS2009,Maciabook2020} and fractal \cite{MandelbrotS1967,Nakayamabook2003}. Fractals can be divided into two categories, i.e., random and deterministic fractals \cite{Mandelbrotbook1977}. The most distinct features of fractal are self-similarity and fractional dimensions. The effective non-integer Hausdorff dimensionality of a fractal is defined by $d_{f}=\text{ln}\,a/\text{ln}\,b$ \cite{HausdorffMA1918,McMullenNMJ1984}, where $a$ denotes how many fractal structures of the previous generation are needed to build the next generation and $b$ represents the ratio of the geometrical size for two adjacent generations. Very recently, the study of the topological phenomena in fractal lattices begins to attract people's attention \cite{SongAPL2014,BrzezinskaPRB2018,PaiPRB2019,FremlingPRR2020,LiuPRL2021,MannaPRB2022,RenNP2023}. Because the fractal lattice lacks translational symmetry and Bloch's theorem is not applicable anymore, the topological invariants thus should be determined in real space \cite{RestaPRL1998,WheelerPRB2019,KangPRB2019}. Owing to the existence of the multiple internal edges and corners, the fractal geometries can support fascinating topologically protected inner edge states and corner states, which have been demonstrated experimentally in acoustic and photonic systems \cite{BiesenthalS2022,ZhengSB2022,LiSB2022}.        

%such as vortex and skyrmion, etc.

%The discovery of higher-order topological insulators (HOTIs) \cite{BenalcazarS2017,SchindlerSA2018,HassanNP2019,XueNM2019,ImhofNP2018,FanPRL2019,Linpj2019} has greatly expanded the scope of topology family \cite{HasanRMP2010,QiRMP2011,Libook2023}, enabling people to have a more comprehensive understanding on topological insulating phases of matter. Generally speaking, the crystal structures are continuous and the translational symmetry holds. The quantized topological invariants in momentum space can therefore be conveniently computed to describe the topological behaviors of topological insulators (TIs), i.e., the so-called ``bulk-boundary" or ``bulk-corner" correspondences \cite{HatsugaiPRL1993,BenalcazarPRB2017,XieNRP2021,LiPR2021}. 
\begin{figure}[!htbp]
\begin{centering}
\includegraphics[width=0.48\textwidth]{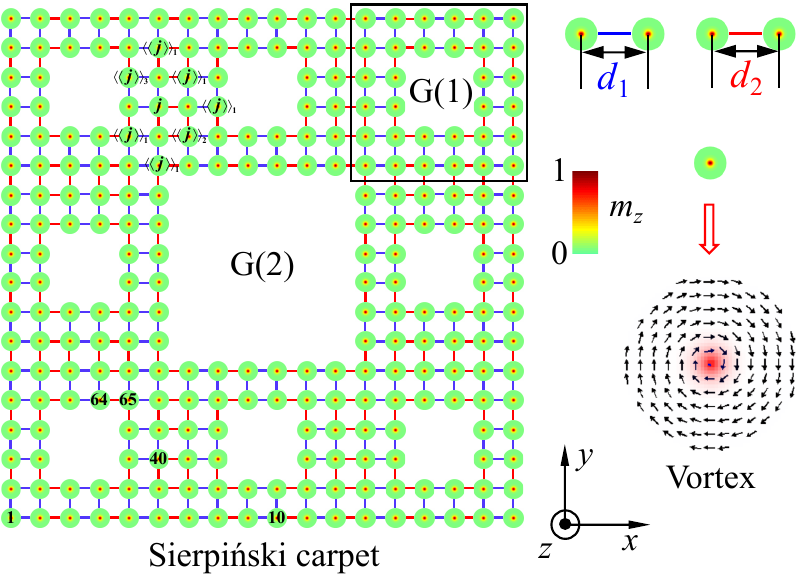}
\par\end{centering}
\caption{Illustration of the Sierpi\'nski carpet array of magnetic vortices. G$(n)(n=1,2,3,...)$ denotes the $n$th generation fractal lattice. $d_{1}$ and $d_{2}$ represent the alternating geometric lengths. The inset plots the micromagnetic structure of a single vortex with topological charge $Q=+1/2$.}
\label{Figure1}
\end{figure}

In magnetic systems, spin waves (or magnons) and magnetic textures (can also be viewed as bound states of infinite magnons) are two important excitations. By utilizing the nonlinear effect, Wu \textit{et al.} \cite{WuPRL2006,RichardsonPRL2018} experimentally observed the spin wave fractal in frequency domain when the power of the input microwave exceeds a critical value. Besides, it has been showed that the spin wave spectra can be tuned over a wide range of frequency in magnetic deterministic fractals, such as Sierpi\'nski carpets \cite{MonceauPLA2010,SwobodaPRB2015} and triangles \cite{ZhouPRB2022,MehtaJPCM2023}. On the other hand, over the past decade, various topological states in magnon- and texture-based crystals with integral dimensions have been reported, for example, the first-order topological insulators (FOTIs) \cite{MookPRB2014,WangPRB2017,LiPRB2018,LiPRB2021}, HOTIs \cite{HirosawaPRL2020,MookPRB2021,LiPRA2020,LiPRB2021-2}, and topological semimetals \cite{FranssonPRB2016,SuPRB2017,MookPRB2017,LiPRA2022}, etc. 
However, the exciting combination of magnetic fractals and topological physics is yet to be explored. One can expect that the magnetic fractal geometry can be used to localize topologically protected spin waves (or texture oscillations), which are particularly helpful for designing magnonic devices with robust multimode transmission channels.     

%the study about the magnetic fractal topology is still lacking. 

\begin{figure*}[ptbh]
\begin{centering}
\includegraphics[width=0.94\textwidth]{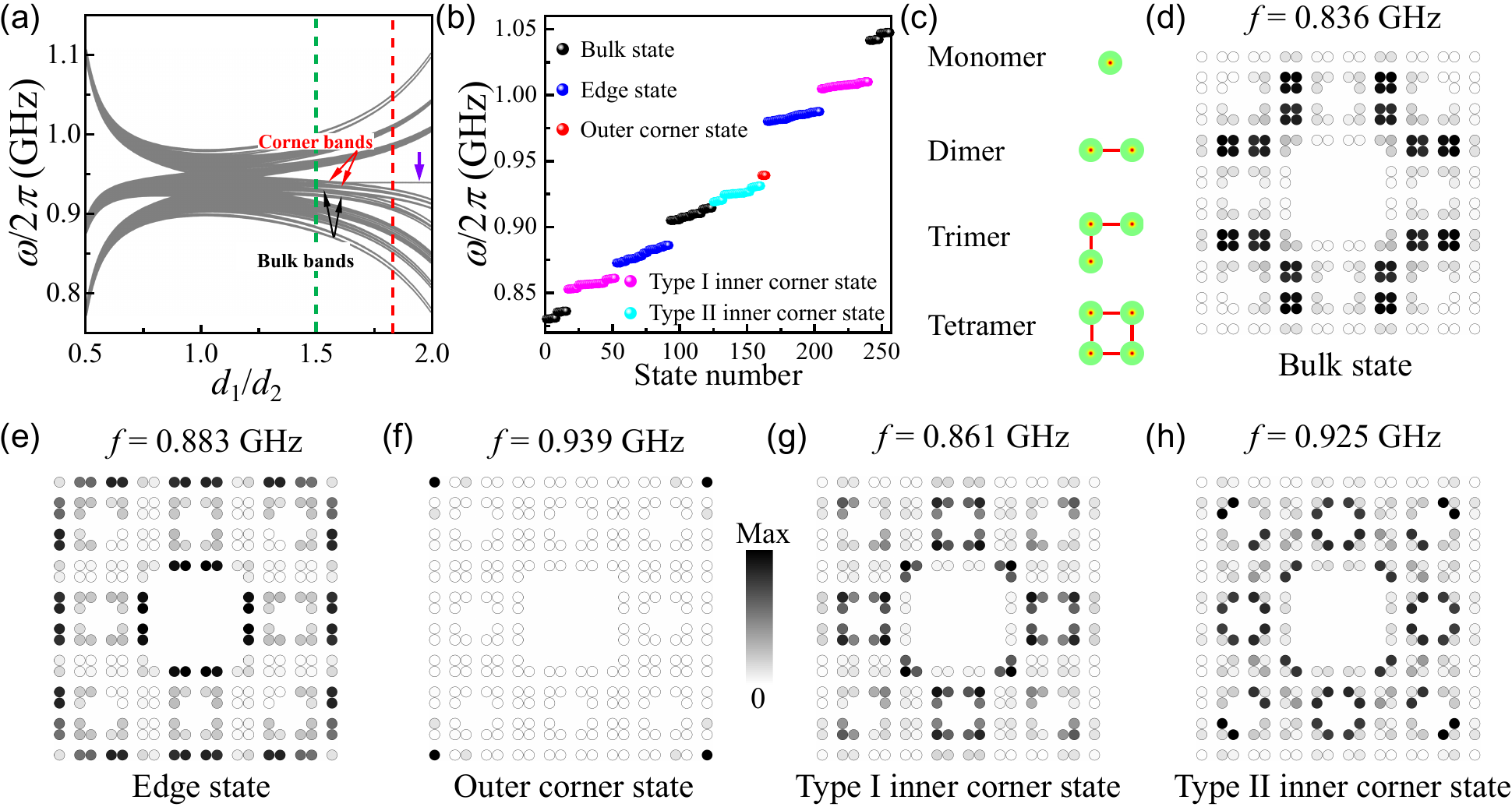}
\par\end{centering}
\caption{(a) The eigenfrequencies of the second generation Sierpi\'nski carpet lattice of magnetic vortices for different values of $d_{1}/d_{2}$. Here we fix $d_{1}+d_{2}=6r$. (b) The eigenfrequencies of the system with $d_{1}/d_{2}=1.83$ as marked by dashed red line in (a). (c) Four different basic elements of Sierpi\'nski carpet lattice when $d_{1}\rightarrow\infty$. The spatial distribution of vortex oscillations for the (d) bulk, (e) edge, (f) outer corner, (g) type I inner corner, and (h) type II inner corner states.}
\label{Figure2}
\end{figure*} 

In this work, we study the topological properties of fractal lattice based on magnetic textures. We take Sierpi\'nski carpet as an example to demonstrate the principle. The vortex is considered as a typical magnetic texture to show the exotic fractal higher-order topological states. By solving Thiele's equation, we obtain the band structure of the magnetic texture fractal lattice. The phase diagram of the system is derived by evaluating the real-space quadrupole moment, from which we conclude that when $d_{1}/d_{2}>1.5$ ($<1.5$), the system is in second-order topological (trivial) phase. Here $d_{1}$ and $d_{2}$ denote the alternating lengths between neighboring vortices, as shown in Fig. \ref{Figure1}. Interestingly, when the system is in the HOTI phase, we discover three different topologically protected corner states with the oscillations confined at outer or inner corners. Our model mimics the two-dimensional Su-Schrieffer-Heeger (SSH) model of higher-order topology as discussed in other systems \cite{LiuPRL2017,KimNPT2020,XiePRB2018}. We also perform micromagnetic simulations to confirm theoretical predictions with good agreement. Our results show that magnetic texture fractal lattices have potential applications for providing abundant topological modes in information processing, which should greatly promote the development of topological spintronics.

\textit{Model}.$-$In deterministic fractals, Sierpi\'nski carpet is one of  the key structures, which has been widely considered for discussing the topological phenomena \cite{RenNP2023,ZhengSB2022,LiSB2022}. In this paper, we focus on the second generation Sierpi\'nski carpet consisting of magnetic vortices, as shown in Fig. \ref{Figure1}. For $n$th generation lattice, the total number of the vortices is $N=4\times8^{n}$. The corresponding Hausdorff dimensionality can be calculated as $d_{f}=\text{ln}\,8/\text{ln}\,3\approx1.893$, which does not depend on the value of the generation. 

The collective dynamics of the vortex lattice can be characterized by the massless Thiele's equation \cite{ThielePRL1973,KimPRL2017}     
\begin{equation}\label{Eq1}
G\hat{z}\times \frac{d\textbf{U}_{j}}{dt}+\textbf{F}_{j}=0.
\end{equation}Here $G = -4\pi$$Qw M_{s}$/$\gamma$ is the gyroscopic coefficient, $Q=\frac{1}{4\pi}\int \!\!\! \int{dxdy\mathbf{m}\cdot(\frac {\partial \mathbf{m}}{\partial {x} } \times \frac {\partial \mathbf{m}}{\partial y } )}$ is the topological charge, $\mathbf {m}$ is the unit vector of the local magnetic moment, $w$ is the thickness of the nanodisk, $M_{s}$ is the saturation magnetization, and $\gamma$ is the gyromagnetic ratio. The displacement of the vortex core from the equilibrium position can be expressed as $\mathbf{U}_{j}= \mathbf R_{j} - \mathbf R_{j}^{0}$. The conservative force $\textbf{F}_{j}=-\partial W / \partial \mathbf U_{j}$ with $W$ being the total potential energy (including the contributions from the confinement of disk boundary and the interaction between nearest-neighbor vortices): $W=\sum_{j}K\textbf{U}_{j}^{2}/2+\sum_{j\neq k}U_{jk}/2$, where $U_{jk}=I_{\parallel}U_{j}^{\parallel}U_{k}^{\parallel}-I_{\perp}U_{j}^{\perp}U_{k}^{\perp}$ \cite{ShibataPRB2003,ShibataPRB2004}. Here $K$ is the spring constant, $I_{\parallel}$ and  $I_{\perp}$ are the longitudinal and transverse coupling constants, respectively. 

Imposing $\textbf{U}_{j}=(u_{j},v_{j})$ and $\psi_{j}=u_{j}+iv_{j}$, and adopting the rotating wave approximation, Eq. \eqref{Eq1} can be recast as \cite{LiPRB2020}  
\begin{equation}\label{Eq2}
  \begin{aligned}
-i\dot{\psi}_{j}=&(\omega_{0}-\frac{\xi^{2}_{1}+\xi^{2}_{2}}{\omega_{0}})\psi_{j}+\sum_{k\in\langle j\rangle,l}\zeta_{l}\psi_{k}-\frac{\xi_{1}\xi_{2}}{2\omega_{0}}\sum_{s\in\langle\langle j\rangle\rangle_{1}}e^{i2\bar{\theta}_{js}}\psi_{s}\\&-\frac{\xi^{2}_{2}}{2\omega_{0}}\sum_{s\in\langle\langle j\rangle\rangle_{2}}e^{i2\bar{\theta}_{js}}\psi_{s}-\frac{\xi^{2}_{1}}{2\omega_{0}}\sum_{s\in\langle\langle j\rangle\rangle_{3}}e^{i2\bar{\theta}_{js}}\psi_{s},
  \end{aligned}
\end{equation}
where ${\omega}_{0}=K/|G|$, $\zeta_{l}=(I_{\parallel, l}-I_{\perp, l})/2|G| $, and $\xi_{l}=(I_{\parallel, l}+I_{\perp, l})/2|G|$, $l=1,2$ denotes the different bond, $\bar{\theta}_{js}$ is the relative geometric angle, and $\langle j\rangle$ is the set of nearest neighbors of $j$. For any site $j$, there exist three different routes to couple next-nearest neighbor sites, i.e., $d_{1}\rightarrow d_{1}$, $d_{2}\rightarrow d_{2}$, and $d_{1}\rightarrow d_{2}$ (or $d_{2}\rightarrow d_{1}$) hoppings, which correspond to $\langle\langle j\rangle\rangle_{3}$, $\langle\langle j\rangle\rangle_{2}$, and $\langle\langle j\rangle\rangle_{1}$, respectively (see Fig. \ref{Figure1}). By solving Eq. \eqref{Eq2}, one can obtain the spectra and eigenmodes of the vortex lattice. In this work, we use the parameters of Permalloy (Py) \cite{YooAPL2012,VeltenAPL2017} nanodisk with thickness $w=10$ nm and radius $r=50$ nm. Then we get the gyrotropic frequency $\omega_{0}=2\pi\times0.939$ GHz, gyroscopic coefficient $G=-3.0725\times10^{-13}$ J\,s\,rad$^{-1}$m$^{-2}$ and spring constant $K=1.8128\times10^{-3}$ J\,m$^{-2}$\cite{Linpj2019}. The explicit expressions for $I_{\parallel}$ and $I_{\perp}$ as a function of $d$ have been determined in our previous work \cite{Linpj2019}: $I_{\parallel}= \mu_{0}M_{s}^{2}r(-1.72064\times10^{-4}+4.13166\times10^{-2}/d^{3}-0.24639/d^{5}+1.21066/d^{7}-1.81836/d^{9})$ and $I_{\perp}=\mu_{0}M_{s}^{2}r(5.43158\times10^{-4}-4.34685\times10^{-2}/d^{3} +1.23778/d^{5}-6.48907/d^{7}+13.6422/d^{9})$, where $d$ is the dimensionless distance parameter normalized by the nanodisk radius $r$ and $\mu_{0}$ is the vacuum permeability.   

\textit{Corner states in fractal lattices}.$-$The spectrum of the second generation Sierpi\'nski carpet (see Fig. \ref{Figure1}) as a function of $d_{1}/d_{2}$ is depicted in Fig. \ref{Figure2}(a). We can see that the system supports several different bands, indicating abundant states emerging in this structure. When $d_{1}/d_{2}$ is greater than a critical value, an isolated band with fixed frequency $\omega/2\pi=0.939$ GHz (gyrotropic frequency of a single vortex) appears, which is the typical feature for corner states [marked by the purple arrow in Fig. \ref{Figure2}(a)]. To analyze localized states in detail, we choose the geometric parameters $d_{1}=194$ nm and $d_{2}=106$ nm ($d_{1}/d_{2}=1.83$), with the eigenfrequencies shown in Fig. \ref{Figure2}(b). By plotting the spatial distribution of the eigenfunctions, we identify five different states: bulk state, edge state, outer corner state, type I inner corner state (with oscillations spreading to all three vortices at the inner corner), and type II inner corner state (only two nonadjacent vortices oscillate), marked by black, blue, red, magenta, and cyan dots in Fig. \ref{Figure2}(b), respectively. The emergence of these states can be intuitively understood by considering the zero-correlation length limit, i.e., $d_{1}\rightarrow\infty$ ($d_{1}=1.83 d_{2}$ and $d_{1}\rightarrow\infty$ are in the same topological phase, as disscussed below). In this case, the system has four different basic elements: monomer, dimer, trimer, and tetramer, as shown in Fig. \ref{Figure2}(c). Therefore, when the vortex oscillations are localized to monomer position, the lattice exhibits the outer corner state [see Fig. \ref{Figure2}(f)]. If the oscillations are confined to dimer and tetramer positions, the system however supports the edge state [see Fig. \ref{Figure2}(e)] and bulk state [see Fig. \ref{Figure2}(d)], respectively. Interestingly, when the trimer positions dominate the oscillations, the vortex lattice show two different inner corner states, i.e., type I [see Fig. \ref{Figure2}(g)] and type II [see Fig. \ref{Figure2}(h)]. This characteristic is reminiscent to the corner states emerging in obtuse-angled corners of breathing honeycomb lattice \cite{LiPRA2020}. 

Since the fractal lattice lacks the translational symmetry, the topological invariant should be calculated in real space to characterize HOTIs. One of the appropriate topological invariant is the real-space quadrupole moment $Q_{xy}$, which is given by \cite{WheelerPRB2019,KangPRB2019,LiSB2022} 
\begin{equation}\label{Eq4}
Q_{xy}=-\frac{i}{2\pi}\text{ln}[\text{det}(S)]\,\,\, \text{mod}\,\,\, 1,\,\,\,\,\,S_{n,m}=V^{*}_{n}\text{exp}\bigg(\frac{i2\pi\hat{x}\hat{y}}{l_{x}l_{y}}\bigg)V_{m},
\end{equation}
\begin{figure}[ptbh]
\begin{centering}
\includegraphics[width=0.48\textwidth]{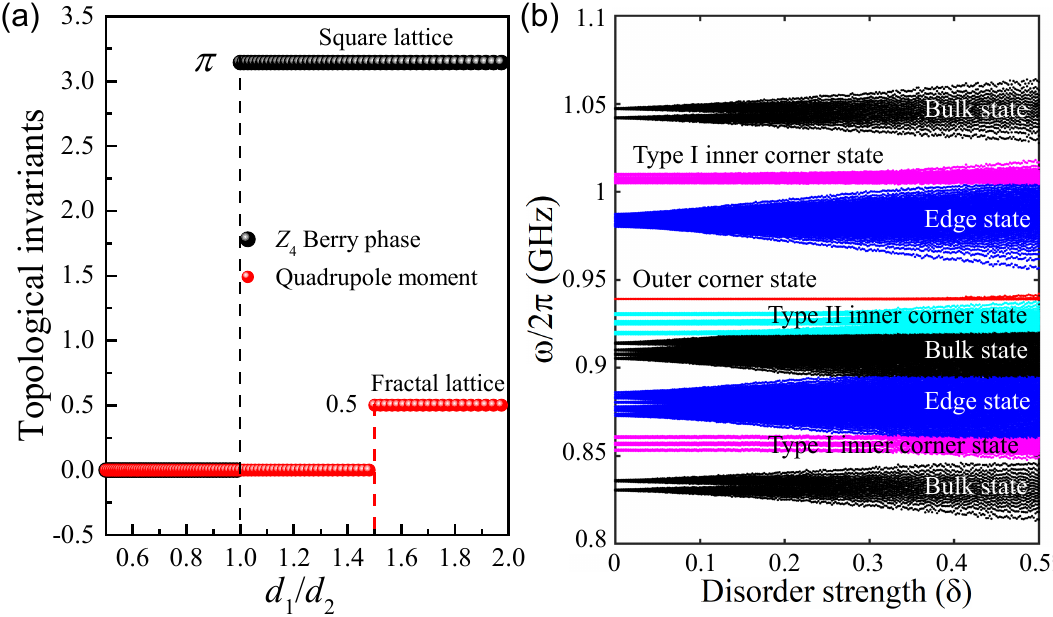}
\par\end{centering}
\caption{(a) Dependence of topological invariants $Z_{4}$ Berry phase (for two-dimensional breathing square lattice) and real-space quadrupole moment (for Sierpi\'nski carpet fractal lattice) on the ratio $d_{1}/d_{2}$. (b) The eigenfrequencies of collective vortex oscillations for fractal lattice under different disorder strengths. Here the value of $d_{1}/d_{2}$ is fixed to 1.83.}
\label{Figure3}
\end{figure}where $V_{n}$ is the $n$th wavefunction [by solving Eq. \eqref{Eq2}] of the fractal lattice with periodic boundary conditions in both $x$ and $y$ directions, $\hat{x}$ ($\hat{y}$) is the position operator along $x$ ($y$) direction, and $l_{x}$ ($l_{y}$) is the length of the lattice in $x$ ($y$) axis. Interestingly, we note that the second generation Sierpi\'nski carpet can be constructed from square lattice by removing nine subsquares (see Fig. \ref{Figure1}). Comparing the band structure of infinite square lattice \cite{LiPRB2020} and Fig. \ref{Figure2}(b), one finds that the corner states of fractal lattice mainly emerge between the third and fourth energy bands. We therefore consider 3/4 band filling (i.e., the bands below the outer corner states are filled) to calculate the real-space quadrupole moment. Figure \ref{Figure3}(a) plots the dependence of $Q_{xy}$ on the ratio $d_{1}/d_{2}$ with red dots. For comparison, the topological invariant $Z_{4}$ Berry phase of square lattice is also plotted with black dots. We can clearly see that $Q_{xy}$ is quantized to 0 when $d_{1}/d_{2}<1.5$ and to 0.5 otherwise, indicating that $d_{1}/d_{2}=1.5$ is the phase transition point separating the trivial and higher-order topological phases. 
\begin{figure*}[ptbh]
\begin{centering}
\includegraphics[width=0.94 \textwidth]{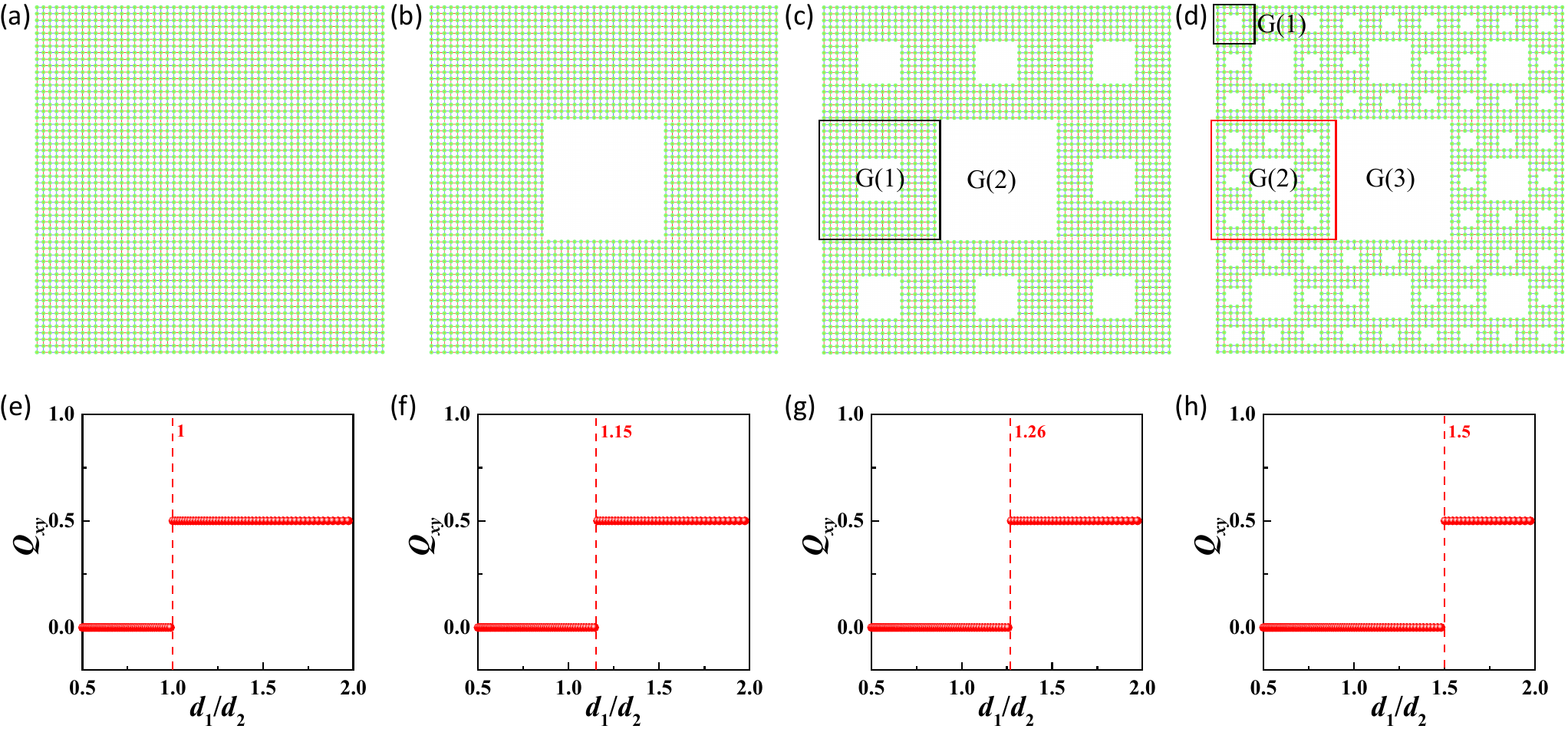}
\par\end{centering}
\caption{(a)-(d) The illustrations of vortex lattices with more and more voids. (e)-(h) The corresponding topological invariant quadrupole moments.}
\label{Figure4}
\end{figure*}Besides, we also identify that the quadrupole moment is origin independent, which is similar to the case in electronic system \cite{MannaPRB2022}. It's well known that the phase transition point is $d_{1}/d_{2}=1$ for square lattice. In such a case, we can say that the nontrivial standard region is squeezed from square lattice [$d_{1}/d_{2}\in(1,+\infty)$] to fractal cases [$d_{1}/d_{2}\in(1.5,+\infty)$]. This ``squeezing" phenomenon can be explained by the spatial translational symmetry breaking resulting from the voids. We further verify that the more severe the spatial translational symmetry breaking, the larger the value of topological phase transition point $d_{1}/d_{2}$, as shown in Fig. \ref{Figure4}. One can naturally expect such a case: if the spatial translational symmetry is totally broken, the system can not support any topological phases, i.e., phase transition point $d_{1}/d_{2}\rightarrow \infty$. Besides, we also find that the phase transition point is independent on the number of the carpet generations [see Figs. \ref{Figure3}(a) and \ref{Figure4}(h)], which can be qualitatively understood by the fact that because of the self-similarity, different generations have the same degree of spatial translational symmetry breaking \cite{LiSB2022}. Moreover, we must point out that the first (second) generation Sierpi\'nski carpets emerging in Figs. \ref{Figure4}(c) and \ref{Figure4}(d) are different due to the distinct duty cycle.

For conventional periodic lattices, at the topological phase transition point, the band gap closing/opening can be clearly observed in momentum space. While we can only obtain the real-space spectrum for fractals owing to the fact that the Bloch's theorem fails to describe the bands. The real-space spectrum exhibits great width for different bands (bulk, edge, and corner). Although it is difficult to identify the gap closing/opening at topological phase transition point, we can find that the corner and bulk [red and black arrows in Fig. \ref{Figure2}(a)] bands begin to separate in frequency when $d_{1}/d_{2}=1.5$, which corresponds to the topological phase transition point.

To further examine whether the corner states emerging in Fig. \ref{Figure2}(b) are topologically protected, we calculate the eigenfrequencies of the fractal vortex lattice under different disorder strengths, with the result being plotted in Fig. \ref{Figure3}(b). Here, the disorders are introduced by supposing that the coupling parameters $\zeta$ and $\xi$ suffer a random change, i.e., $\zeta \rightarrow \zeta(1+\delta Z)$ and $\xi \rightarrow \xi(1+\delta Z)$, where $\delta$ denotes the strength of the disorder and $Z$ is a random number uniformly distributed between $-1$ and $1$, which apply to all vortices. We average the numerical results for 100 realizations to avoid the error from single calculation. From Fig. \ref{Figure3}(b), we can see that both the frequencies of outer and inner corner states are robust for moderate disorder strength. Interestingly, we find the critical value of disorder strength for shifting the frequencies of outer corner states is larger than those for inner corner states, which indicates that the outer corner state has stronger topological stability compared to inner ones. 

Besides, if we introduce random empty sites in the square lattice, localized states around the empty sites may possibly appear. However, these local modes are not protected by symmetry and therefore are trivial. As a result, the critical difference between a fractal and a conventional square lattice with random empty sites is that the corner states are topologically protected in fractals while not in conventional square lattice with random empty sites.

%\textcolor{RED}{One the one hand, our model has some similar characteristics compared to tight-binding Hamiltonian. At first, the mechanism for generating higher-order topology is the same, i.e., the two-dimensional SSH model \cite{LiuPRL2017,KimNPT2020,XiePRB2018}. Secondly, both the higher-order topological phases are protected by $C_{4}$ symmetry. On the other hand, there exist key differences between our model and tight-binding model. Our model contains the extra on-site terms and second-nearest hopping terms, which break the chiral symmetry. As a result, the corner states and bulk states are spectrally separated in our model, while for the tight-binding fractal system, these states spectrally overlap with each other \cite{ZhengSB2022}. Remarkably, this feature provide advantages to distinguish and detect different modes.}
Our model can also be mapped to the tight-binding model with the following Hamiltonian
\begin{equation}\label{Eq5}
  \begin{aligned}
\mathcal{H}&=\sum_{j}(\omega_{0}-\frac{\xi^{2}_{1}+\xi^{2}_{2}}{\omega_{0}})\psi_{j}^{*}\psi_{j}+\zeta\sum_{\langle jk\rangle}\psi_{j}^{*}\psi_{k}-\frac{\xi_{1}\xi_{2}}{2\omega_{0}}\sum_{\langle\langle jk\rangle\rangle_{1}}e^{i2\bar{\theta}_{jk}}\psi_{j}^{*}\psi_{k}\\&-\frac{\xi_{2}^{2}}{2\omega_{0}}\sum_{\langle\langle jk\rangle\rangle_{2}}e^{i2\bar{\theta}_{jk}}\psi_{j}^{*}\psi_{k}-\frac{\xi_{1}^{2}}{2\omega_{0}}\sum_{\langle\langle jk\rangle\rangle_{3}}e^{i2\bar{\theta}_{jk}}\psi_{j}^{*}\psi_{k}+c.c.
  \end{aligned}
\end{equation}
Comparing to the two-dimensional SSH model \cite{LiuPRL2017,KimNPT2020,XiePRB2018}, our Hamiltonian contains an extra on-site term $\sum_{j}[\omega_{0}-(\xi^{2}_{1}+\xi^{2}_{2})/\omega_{0}]\psi_{j}^{*}\psi_{j}$, which breaks the chiral symmetry. As a result, the corner states and bulk states are spectrally separated in our model, while for the SSH system, these states spectrally overlap with each other. Remarkably, this feature provide advantages to distinguish different modes in our system. Besides, the mechanism for generating higher-order topology is analogue to SSH model. The higher-order topological states emerging in our model are protected by $C_{4}$ symmetry.

\textit{Micromagnetic simulations}.$-$To verify the theoretical predictions about the fractal higher-order topological states, we perform full micromagnetic simulations \cite{Footnote2}. The second generation Sierpi\'nski carpet array consisting of 256 identical magnetic vortices is considered, as shown in Fig. \ref{Figure1}. 
\begin{figure}[ptbh]
\begin{centering}
\includegraphics[width=0.46\textwidth]{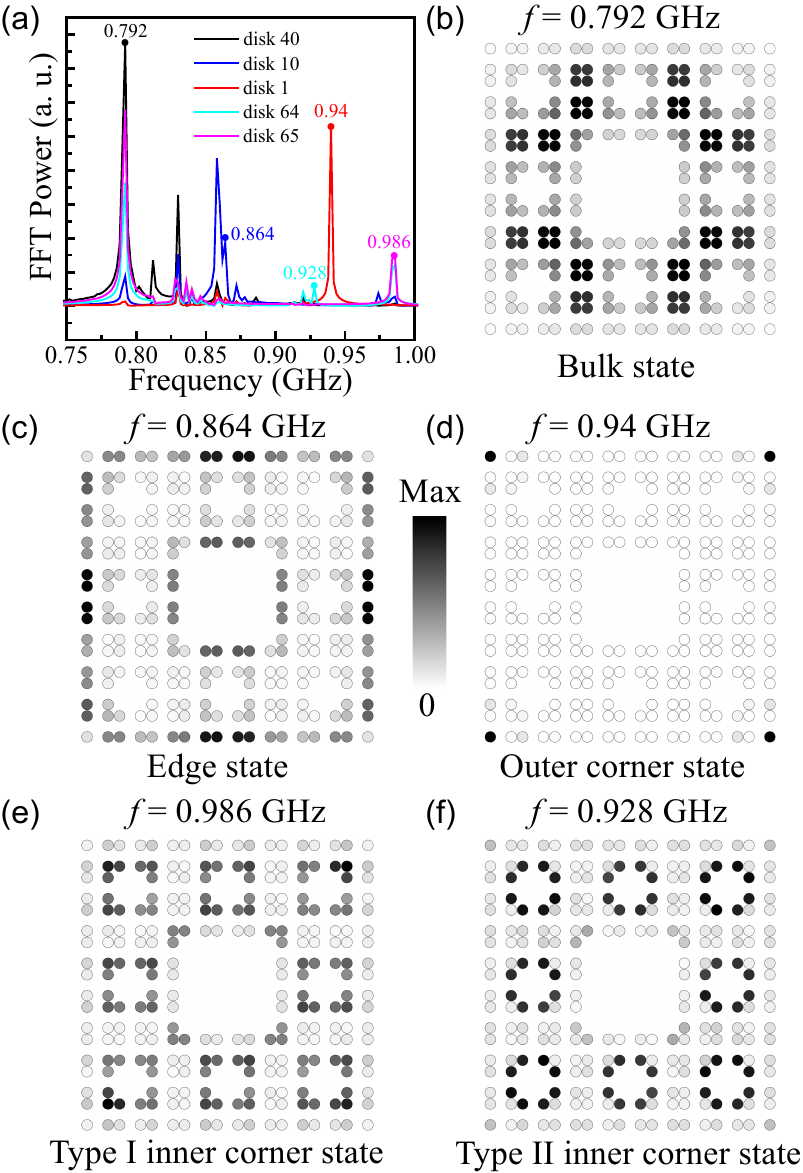}
\par\end{centering}
\caption{(a) The temporal Fourier spectra of the vortex oscillations at different positions as marked in Fig. \ref{Figure1}. (b)-(f) The spatial distribution of FFT intensity for different modes.}
\label{Figure5}
\end{figure}
We compute the temporal fast Fourier transform (FFT) spectrum of the vortex oscillations at different positions, labeled by arabic numbers 1, 10, 40, 64, and 65 in Fig. \ref{Figure1}. Figure \ref{Figure5}(a) plots the numerical results, with black, blue, red, cyan, and magenta curves representing the positions of bulk (No. 40), edge (No. 10), outer corner (No. 1), inner corner 1 (No. 64), and inner corner 2 (No. 65) bands, respectively. From Fig. \ref{Figure5}(a), we find that near the frequency of 0.94 GHz, the outer corner band has a very strong peak, while other bands do not, which is an obvious feature of outer corner state with oscillations confined only at four outer corners. The frequency range supporting other states can be obtained through the similar principle. The spatial distribution of the FFT intensity with representative frequencies are plotted in Figs. \ref{Figure5}(b)-\ref{Figure5}(f) to visualize the characteristic of different modes. We can identify the bulk state, edge state, outer corner state, and inner corner states of both type I and II with vortex oscillation localized at tetramer [see Fig. \ref{Figure5}(b)], dimer [see Fig. \ref{Figure5}(c)], monomer [see Fig. \ref{Figure5}(d)], and trimer [see Figs. \ref{Figure5}(e) and \ref{Figure5}(f)], respectively. Interestingly, due to the absence of dimers in the smallest vacant squares, the type I inner corner states exhibit the exotic characteristics with oscillation spreading over all vortices in these regions. These results agree well with the theoretical calculations presented in Fig. \ref{Figure2}.    

\textit{Discussion and conclusion}.$-$From an experimental perspective, the fabrication and detection of HOTIs in magnetic texture fractal are fully within the reach of current technology. On the one hand, the artificial fractal of magnetic nanodisks (it is convenient to obtain vortex or skyrmion states when appropriate parameters are chosen) can be created with electron-beam lithography \cite{BehnckePRB2015,SunPRL2013} or X-ray illumination \cite{GuangNC2020}. On the other hand, the nanometer-scale magnetic texture positions can be tracked by using a biased conductive scanning nanoscale tip \cite{YuPRA2021} or ultrafast Lorentz microscopy technique \cite{MollerCP2020}. The magnetic texture arrays thus provide an ideal platform for studying fractal topology. In addition, higher-order topological states emerging in magnetic texture fractal lattices are expected to have a lot of potential applications for information processing. For example, the topologically protected inner corner states in fractal geometry provides massive oscillation sources, which can be used to design multifrequency robust nano-oscillators. Besides, by constructing fractal lattice with different generation, one can achieve the localization of magnetic texture oscillations (information) at desired positions for display application.          

Comparing to two-dimensional square lattice with translational symmetry, our model exhibits several unusual properties, apart from some similarities. Firstly, the Sierpi\'nski carpet is constructed by cutting numerous subsquares from the two-dimensional square lattice. Therefore, the unit cell contains four sites both for fractal and square lattices. Secondly, due to the existence of multiple internal edges and corners, the fractal lattices have more edge and corner states than conventional square lattices. Thirdly, when considering the second-order topological phase, there only exist three basic elements for square lattices, i.e., monomer, dimer, and tetramer, however, the fractal lattice has additional trimer. As a result, the fractal lattices can support exotic inner corner states, while square lattices cannot. At last, both the corner states emerging in fractal and square lattices are protected by the fourfold rotation symmetry.

In this work, we only focus on the Sierpi\'nski carpet array, while other fractal lattices, for example, Sierpi\'nski gasket \cite{RenNP2023,BiesenthalS2022} and  Koch curve \cite{SongAPL2014} should support exotic localized topological modes too. Besides the magnetic vortex, there exist several other magnetic textures, such as skyrmion \cite{MuhlbauerS2009,JiangS2015}, magnetic bubble \cite{MoonPRB2014,MakhfudzPRL2012}, and domain wall \cite{AtkinsonNM2003,CatalanRMP2012} \sout{etc}. The study about the fractal TIs based on these magnetic textures is an appealing research topic, from which one can expect the multichannel propagation of magnons and topologically protected multimode oscillators. To accurately describe the dynamics of vortex, the higher-order terms, like the mass term and non-Newtonian gyration term shoud be added in Thiele's equation \cite{Linpj2019,LiPRB2018}. In such a case, the system then may provide more topologically protected outer and inner corner states, which is also an interesting issue for study.    

To summarize, we have investigated the higher-order topological phases in second generation Sierpi\'nski carpet of magnetic vortices. The band structures of the collective vortex oscillations were calculated by solving Thiele's equation. We found that the fractal lattice can exhibit one type of outer corner state and two different types of inner corner states under a proper geometric condition. By evaluating the real-space quadrupole moment as the topological invariant, we obtain the full phase diagram. We showed that both outer and inner corner states are topologically robust against moderate disorder. Full micromagnetic simulations were performed to verify the theoretical predictions with a great agreement. Our findings provide important theoretical reference to investigating the in-gap states in magnetic texture based fractals, which also represent a crucial step for combining fractal and topological physics. 

\begin{acknowledgments}
\textit{Acknowledgments}.$-$We thank Z. Wang and X. S. Wang for helpful discussions. This work was supported by the National Key R\&D Program under Contract No. 2022YFA1402802 and the National Natural Science Foundation of China (NSFC) (Grants No. 12074057 and No. 12374103). Z.-X.L. acknowledges financial support from the NSFC (Grant No. 11904048) and the Natural Science Foundation of Hunan Province of China (Grant No. 2023JJ40694).  
\end{acknowledgments}

\end{document}